\documentclass[prl,twocolumn,amsmath,superscriptaddress,amssymb]{revtex4}

\usepackage{graphicx}
\usepackage{dcolumn}
\usepackage{bm}
\usepackage{amssymb}
\usepackage{amsmath}
\usepackage{wasysym}
\usepackage{color}
\usepackage{times}

\usepackage{placeins}
\usepackage{epstopdf}
\usepackage{lipsum}
\usepackage[separate-uncertainty=true]{siunitx}

\makeatletter
\renewcommand*\env@matrix[1][c]{\hskip -\arraycolsep
  \let\@ifnextchar\new@ifnextchar
  \array{*\c@MaxMatrixCols #1}}
\makeatother
\begin{document}

\title{Two-dimensional ferromagnetic extension of a topological insulator}


\author{P. Kagerer}\affiliation{Physikalisches Institut (EP7), Universität Würzburg, Am Hubland, D-97074 W\"urzburg,  Germany}\affiliation{W\"urzburg-Dresden Cluster of Excellence {\it ct.qmat}, Germany}

\author{C. I. Fornari}\email{celso.fornari@physik.uni-wuerzburg.de}\affiliation{Physikalisches Institut (EP7), Universität Würzburg, Am Hubland, D-97074 W\"urzburg,  Germany}\affiliation{W\"urzburg-Dresden Cluster of Excellence {\it ct.qmat}, Germany}

\author{S. Buchberger}\affiliation{Physikalisches Institut (EP7), Universität Würzburg, Am Hubland, D-97074 W\"urzburg,  Germany}\affiliation{W\"urzburg-Dresden Cluster of Excellence {\it ct.qmat}, Germany}

\author{T. Tschirner}\affiliation{W\"urzburg-Dresden Cluster of Excellence {\it ct.qmat}, Germany}\affiliation{Leibniz IFW Dresden, Helmholtzstr. 20,
01069 Dresden, Germany}

\author{L. Veyrat}\affiliation{W\"urzburg-Dresden Cluster of Excellence {\it ct.qmat}, Germany}\affiliation{Leibniz IFW Dresden, Helmholtzstr. 20,
01069 Dresden, Germany}\affiliation{Physikalisches Institut (EP4), Universität Würzburg, Am Hubland, D-97074 W\"urzburg, Germany}

\author{M. Kamp}\affiliation{Physikalisches Institut and R\"ontgen-Center for Complex Material Systems (RCCM),
Fakult\"at f\"ur Physik und Astronomie, Universit\"at W\"urzburg, Am Hubland, D-97074 W\"urzburg, Germany}

\author{A. V.  Tcakaev}\affiliation{W\"urzburg-Dresden Cluster of Excellence {\it ct.qmat}, Germany}\affiliation{Physikalisches Institut (EP4), Universität Würzburg, Am Hubland, D-97074 W\"urzburg, Germany}

\author{V. Zabolotnyy}\affiliation{W\"urzburg-Dresden Cluster of Excellence {\it ct.qmat}, Germany}\affiliation{Physikalisches Institut (EP4), Universität Würzburg, Am Hubland, D-97074 W\"urzburg, Germany}

\author{S. L. Morelh\~ao}\affiliation{Instituto de F\'isica, Universidade de S\~ao Paulo, 05508-090 S\~ao Paulo, SP, Brazil}

\author{B. Geldiyev}\affiliation{Physikalisches Institut (EP7), Universität Würzburg, Am Hubland, D-97074 W\"urzburg,  Germany}\affiliation{W\"urzburg-Dresden Cluster of Excellence {\it ct.qmat}, Germany}

\author{S. Müller}\affiliation{Physikalisches Institut (EP7), Universität Würzburg, Am Hubland, D-97074 W\"urzburg,  Germany}\affiliation{W\"urzburg-Dresden Cluster of Excellence {\it ct.qmat}, Germany}

\author{A. Fedorov}\affiliation{W\"urzburg-Dresden Cluster of Excellence {\it ct.qmat}, Germany} \affiliation{Leibniz IFW Dresden, Helmholtzstr. 20,
01069 Dresden, Germany} \affiliation{Helmholtz-Zentrum Berlin f\"ur Materialien und Energie, Albert-Einstein-Str. 15, D-12489 Berlin, Germany}

\author{E. Rienks} \affiliation{Helmholtz-Zentrum Berlin f\"ur Materialien und Energie, Albert-Einstein-Str. 15, D-12489 Berlin, Germany}

\author{P. Gargiani}\affiliation{ALBA Synchrotron Light Source, E-08290 Cerdanyola del Valles, Spain}

\author{M. Valvidares}\affiliation{ALBA Synchrotron Light Source, E-08290 Cerdanyola del Valles, Spain}

\author{L. C. Folkers}\affiliation{W\"urzburg-Dresden Cluster of Excellence {\it ct.qmat}, Germany}\affiliation{Institut f\"ur Festk\"orper- und Materialphysik, Technische Universit\"at Dresden, D-01062 Dresden, Germany}

\author{A. Isaeva}\affiliation{Leibniz IFW Dresden, Helmholtzstr. 20,
01069 Dresden, Germany}\affiliation{Van der Waals – Zeeman Institute, IoP, University of Amsterdam, 1098 XH Amsterdam, The Netherlands}

\author{B. Büchner}\affiliation{W\"urzburg-Dresden Cluster of Excellence {\it ct.qmat}, Germany}\affiliation{Leibniz IFW Dresden, Helmholtzstr. 20,
01069 Dresden, Germany}

\author{V. Hinkov}\affiliation{W\"urzburg-Dresden Cluster of Excellence {\it ct.qmat}, Germany}\affiliation{Physikalisches Institut (EP4), Universität Würzburg, Am Hubland, D-97074 W\"urzburg, Germany}

\author{R. Claessen}\affiliation{W\"urzburg-Dresden Cluster of Excellence {\it ct.qmat}, Germany}\affiliation{Physikalisches Institut (EP4), Universität Würzburg, Am Hubland, D-97074 W\"urzburg, Germany}

\author{H. Bentmann}\email{Hendrik.Bentmann@physik.uni-wuerzburg.de}\affiliation{Physikalisches Institut (EP7), Universität Würzburg, Am Hubland, D-97074 W\"urzburg, Germany}\affiliation{W\"urzburg-Dresden Cluster of Excellence {\it ct.qmat}, Germany}

\author{F. Reinert}\affiliation{Physikalisches Institut (EP7), Universität Würzburg, Am Hubland, D-97074 W\"urzburg, Germany}\affiliation{W\"urzburg-Dresden Cluster of Excellence {\it ct.qmat}, Germany}

\date{\today}

\begin{abstract}
Inducing a magnetic gap at the Dirac point of the topological surface state (TSS) in a 3D topological insulator (TI) is a route to dissipationless charge and spin currents. Ideally, magnetic order is present only at the surface and not in the bulk, e.g. through proximity of a ferromagnetic (FM) layer. However, such a proximity-induced Dirac mass gap has not been observed, likely due to insufficient overlap of TSS and the FM subsystem. Here, we take a different approach, namely FM extension, using a thin film of the 3D TI Bi$_2$Te$_3$, interfaced with a monolayer of the lattice-matched van der Waals ferromagnet MnBi$_2$Te$_4$. Robust 2D ferromagnetism with out-of-plane anisotropy and a critical temperature of $\text{T}_\text{c}\approx$~15 K is demonstrated by X-ray magnetic dichroism and electrical transport measurements. Using angle-resolved photoelectron spectroscopy, we observe the opening of a sizable magnetic gap in the 2D FM phase, while the surface remains gapless in the paramagnetic phase above T$_c$. This sizable gap indicates a relocation of the TSS to the FM ordered Mn moments near the surface, which leads to a large mutual overlap. 
\end{abstract}
\maketitle

Engineering the surface of a topological insulator (TI) to host ferromagnetism is expected to enable unconventional phenomena, including topological magneto-electric effects and Majorana-fermion quasiparticles, with potential applications ranging from spintronics to quantum computation \cite{bernevig2022progress,Chang2013:QAH_dopedTI,Tokura2019:MTI}. A paradigmatic scenario to achieve this is to interface the TI with a ferromagnetic (FM) layer, aiming to induce an exchange gap in the surface Dirac cone through magnetic proximity while preserving the bulk topology \cite{Zhang:11.10}. However, although various FM-TI heterostructures have been investigated \cite{wray2011topological,PhysRevLett.108.256810,lang2014proximity,tang2017above,katmis2016high,hirahara:20_fabricationMonolayer,lee2016direct,PhysRevLett.123.016804}, the clear observation of magnetic topological behaviour in such systems remains elusive. In particular, no direct measurement of a magnetic gap in the surface state has been reported so far. These difficulties likely arise from a weak hybridization at the FM-TI interface \cite{Tokura2019:MTI,PhysRevLett.125.226801,PhysRevLett.128.126802,eremeev2015interface}, inhibiting sizable effects on the topological surface state (TSS). 

\begin{figure*}[t]
\includegraphics[width=.95\textwidth]{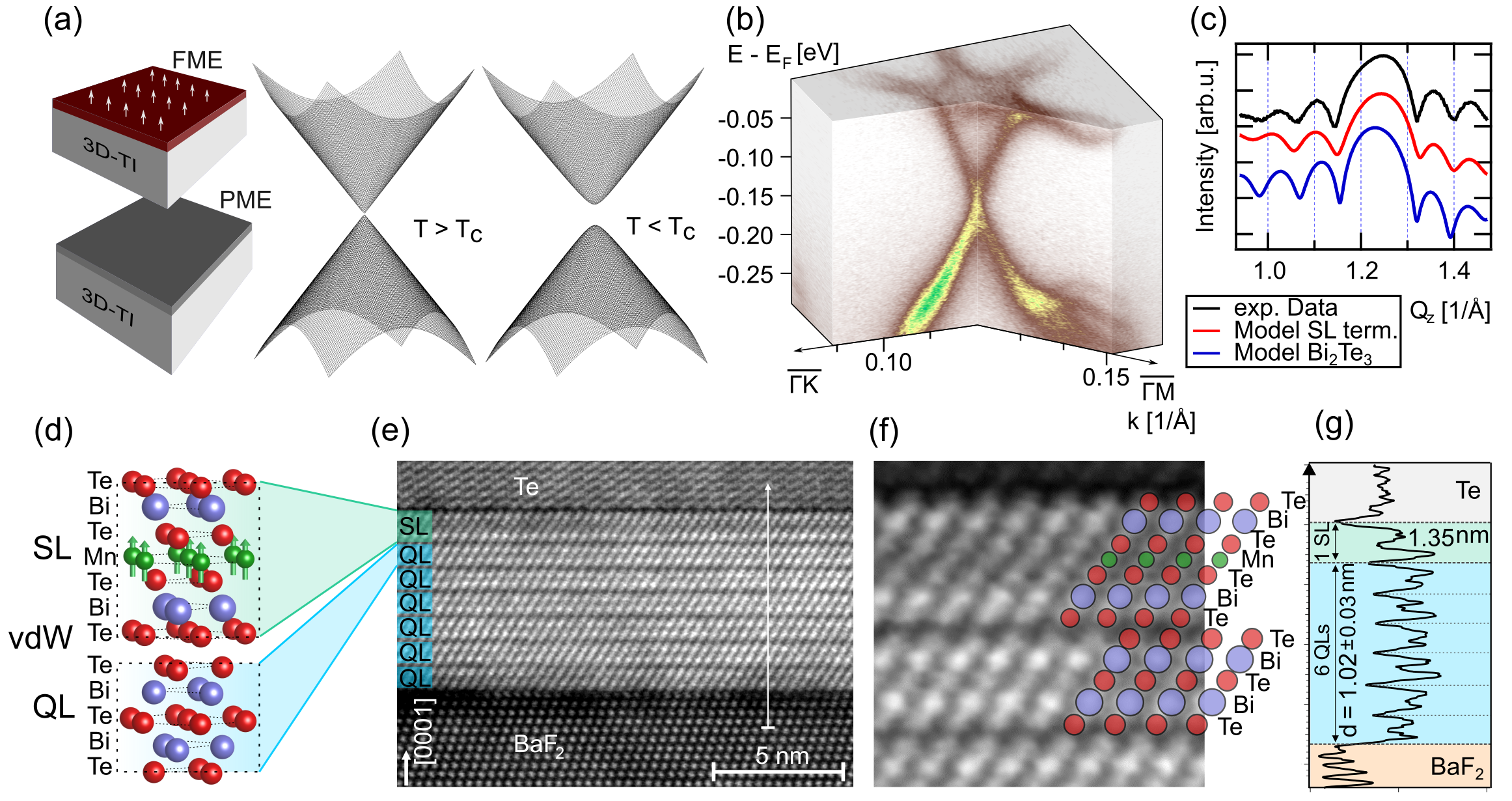}
\caption{(a) Scheme of a magnetically extended 3D TI in the ferro- (FME) and paramagnetic (PME) cases. Below T$_c$, ferromagnetic order in out-of-plane direction induces a magnetic gap in the topological surface state. (b) 3D ARPES data set of the topological surface state, measured above $T_c$ at $T =$~20~K using p-polarized light at h$\nu$ = \SI{12}{\electronvolt}. (c) X-ray diffraction-Q$_z$-scan around the (0006) Bragg peak of Bi$_2$Te$_3$. The line-shape asymmetry is attributed to the single MnBi$_2$Te$_4$ layer, as confirmed by calculations. (d) Schematic of the layer structure. The 3D TI Bi$_2$Te$_3$ is terminated by a single layer of MnBi$_2$Te$_4$ containing the magnetically active Mn atomic sheet. (e,f) Cross-sectional STEM image revealing the sample layer stacking: BaF$_2$ substrate, 6 quintuple-layer Bi$_2$Te$_3$ film, single septuple-layer (SL) MnBi$_2$Te$_4$ and Te-capping layer. (g) Intensity profile analysis along the trace shown in (e).}
\label{StructureFigure}
\end{figure*}

At the same time, it has been proposed that suitable van der Waals (vdW) heterostructures, with weak potential modulation at the interface, may allow the TSS wave function to relocate from the TI surface into the adjacent magnetic layer. Such a ferromagnetic extension (FME) is expected to dramatically enhance the magnetic gap compared to a mere proximity effect \cite{otrokov2017highly}.  In this work we explore the magnetic and topological properties of an epitaxial vdW heterostructure consisting of the TI Bi$_2$Te$_3$ and a monolayer MnBi$_2$Te$_4$ \cite{otrokov:19}. The structural and chemical similarity of the two compounds has been predicted to generate a FME of Bi$_2$Te$_3$ \cite{otrokov2017highly}. As schematically shown in Fig.~\ref{StructureFigure}(a), the FME is expected to break global time-reversal symmetry (TRS) at the surface, allowing for the opening of a magnetic gap in the TSS below $T_c$ \cite{Zhang:11.10}. 
However, previous experimental attempts to realize a FME could not demonstrate a correlation between electronic structure and magnetic order \cite{hirahara2017large,hirahara:20_fabricationMonolayer,fukusawa:21_absenceFMmonolayer,li:2021designerferromagnet}.





We used molecular beam epitaxy to grow a single monolayer MnBi$_2$Te$_4$ on an epitaxial Bi$_2$Te$_3$ epilayer layer, as confirmed by high-resolution X-ray diffraction (XRD) and scanning transmission electron microscopy (STEM). X-ray absorption (XAS) and x-ray magnetic circular dichroism (XMCD) were measured at the BOREAS beamline at the ALBA synchrotron in Barcelona. Angle resolved photoemission spectroscopy (ARPES) measurements were performed at the One-Cube endstation at the BESSY II synchrotron in Berlin. Details of the film growth and characterization methods are presented in the supplement. Fig.~\ref{StructureFigure}(e) shows a cross-section STEM image of a heterostructure. The films are grown on insulating (111) BaF$_2$ substrates and capped with a Te protective layer to avoid surface oxidation. VdW gaps parallel to the surface are clearly observed, evidencing the formation of high quality layers and the absence of twinned domains [Fig. (S4)]. A single MnBi$_2$Te$_4$ SL is present on the surface, as verified by the zoom-in STEM image in Fig.~\ref{StructureFigure}(f). The difference in contrast allows to distinguish the three elements on their respective sites. Figure~\ref{StructureFigure}(g) shows a line profile extracted along the [0001] direction [white arrow in Fig.~\ref{StructureFigure}(e)]. We obtain an average thickness of 1.02(3) nm for the QLs and of 1.35 nm for the SL, consistent with values reported for the individual compounds \cite{zeugner:2019MBTStructural,Morelhao:17XRDmodel}. In Fig.~\ref{StructureFigure}(c) we compare an experimental Q$_z$-XRD-scan, recorded around the Bi$_2$Te$_3$ (0006) Bragg peak, with simulated scans for Bi$_2$Te$_3$ with and without an additional SL of MnBi$_2$Te$_4$. The experimental line shape shows a pronounced asymmetry with a shoulder at lower Q$_z$, which, based on our XRD simulations [see SI Fig. (S5) for details], can be assigned to the presence of a single MnBi$_2$Te$_4$ SL.

Fig.~\ref{StructureFigure}(b) shows ARPES data near the Fermi level (E$_F$) along high-symmetry directions. The data were obtained in the paramagnetic regime of the MnBi$_2$Te$_4$ layer above T$_c$ (cf. Figs.~2 and 3). A TSS with Dirac-like band dispersion is observed, as further confirmed by the photon-energy-dependent data in Fig.~(S9). Interestingly, the TSS dispersion is strongly modified from the one of a pristine Bi$_2$Te$_3$(0001) surface \cite{chen:09}. In agreement with theoretical calculations \cite{otrokov2017highly}, the Dirac point is shifted upwards in energy into the bulk band gap. As a result, both, the upper and lower part of the Dirac cone display a linear dispersion near the $\bar{\Gamma}$ point, in contrast to Bi$_2$Te$_3$(0001). Moreover, the Fermi surface acquires a pronounced hexagonal star-shape with cusps along $\overline{\Gamma M}$, in accordance with recent calculations \cite{li:2021designerferromagnet}. The strong effect of the MnBi$_2$Te$_4$ single layer on the TSS band dispersion and the overall agreement with calculations support a scenario where the TSS wave function relocates into the SL, as predicted theoretically \cite{otrokov2017highly}.    
\\

\begin{figure*}[htb]
\includegraphics[width=.95\textwidth]{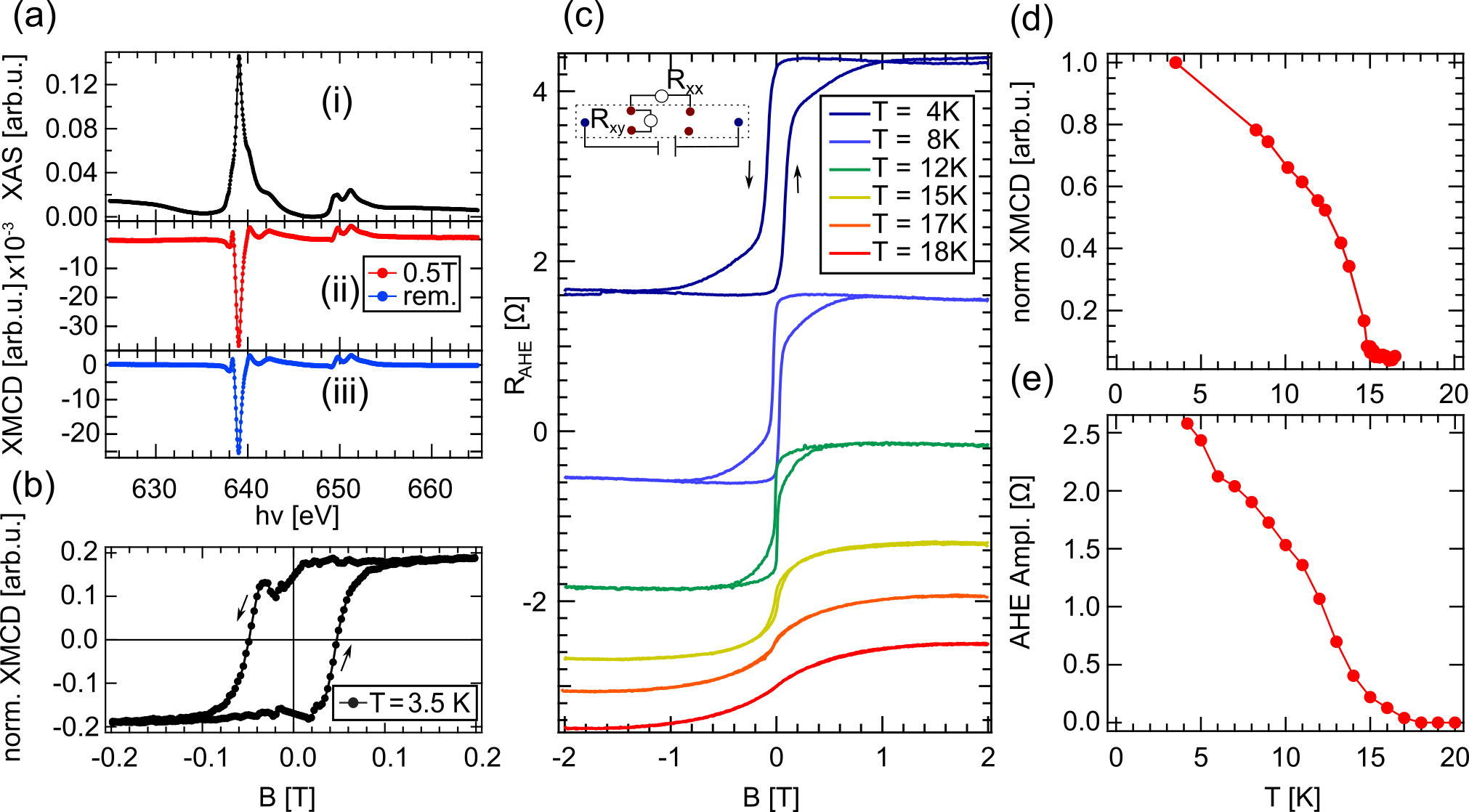}
\caption{(a) XAS and XMCD data confirming a stable ferromagnetic state of MnBi$_2$Te$_4$. (i) XAS data set at the Mn L$_{2,3}$-edge. (ii,iii) XMCD difference spectra at $T=$~3.5~K for a saturation field of 0.5~T and in remanence. (b) XMCD hysteresis at the same temperature indicating long-range ferromagnetic order. (c) Anomalous Hall effect (AHE) in a single-layer MnBi$_2$Te$_4$/Bi$_2$Te$_3$ heterostructure versus magnetic field, at several temperatures. The arrows indicate the magnetic field sweep direction. Curves are vertically shifted for clarity. A clear hysteretic behaviour of the AHE is observed below \SI{18}{\kelvin}, with coercive fields up to \SI{90}{\milli\tesla}, confirming the ferromagnetic behaviour of the single MnBi$_2$Te$_4$ layer. The inset shows a schematic of the pseudo-Hall bar geometry of the transport experiments.(d) Temperature dependence of the remnant XMCD signal yielding a critical temperature of 14.9~K. (e) Temperature dependence of the AHE amplitude at zero field extracted from the data in (c).
}
\label{Magnetism_figure}
\end{figure*}


\begin{figure*}[htb]
\includegraphics[width=1\textwidth]{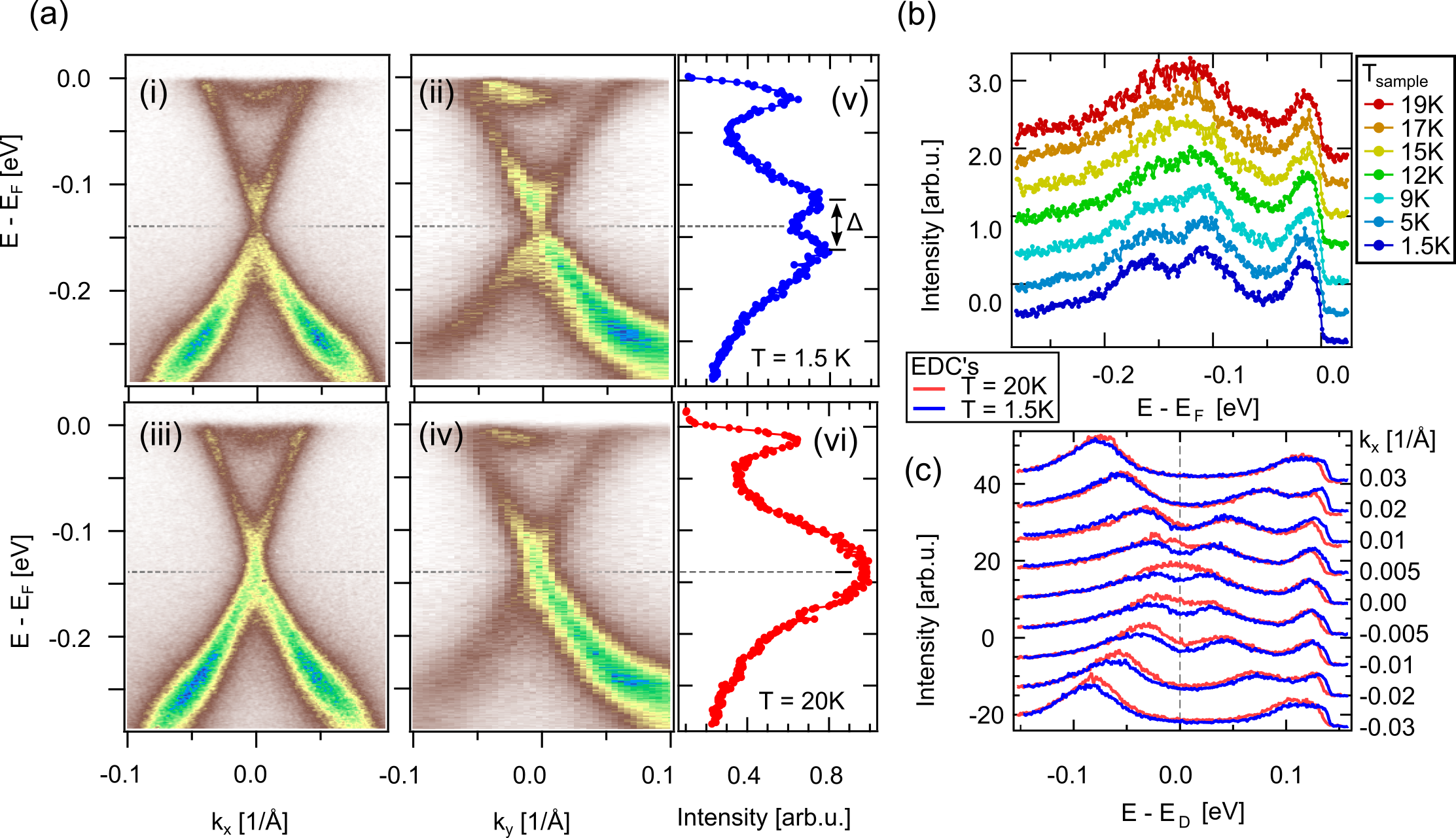}
\caption{(a) ARPES data acquired below (top-row) and above (bottom row) the critical temperature T$_c$ using p-polarized light at $h\nu = $~12~eV. Data sets acquired perpendicular [panels (i) and (iii)] and along [panels (ii) and (iv)] the plane of incidence are shown. Panels (v,vi) display energy distribution curves (EDC) at the the $\bar{\Gamma}$-point, revealing a magnetic exchange splitting at the Dirac point below T$_c$. (b) Temperature-dependent EDC at the $\bar{\Gamma}$-point, showing the spectral-weight evolution across T$_c$. (c) Momentum-dependent EDC above and below $T_c$, taken from the data sets in (a). The energy scales are relative to the respective Dirac-point position, to compensate for a small temperature-dependent energy shift. The ferromagnetic order affects the spectral weight predominantly in a narrow momentum range of $\delta k=\pm$0.01\AA$^{-1}$ around the $\bar{\Gamma}$-point.}
\label{ARPES_Figure}
\end{figure*}


We now demonstrate the presence of ferromagnetism in the MnBi$_2$Te$_4$ monolayer, employing X-ray magnetic circular dichroism (XMCD) and electrical transport. The data was collected in total-electron-yield mode with light incidence and external field oriented along the surface normal. Fig.~\ref{Magnetism_figure}(a) shows X-ray absorption (XAS) and XMCD data in saturation and in remanence. The XAS lineshape of the Mn \textit{L}$_{2,3}$ absorption edge strongly resembles the one of bulk MnBi$_2$Te$_4$ \cite{otrokov:19}, indicating Mn ions in the same oxidation state. The apparent absence of Mn oxide components in the spectra further verifies the successful preparation of pristine surfaces through mechanical removal of the Te cap, as established earlier for Bi$_2$Te$_3$ \cite{fornari:16_capping}. As seen from the XMCD signals in saturation and remanence, the MnBi$_2$Te$_4$ monolayer hosts a stable ferromagnetic state at the lowest measured temperature of $\sim$\SI{3.5}{\kelvin}, with a remanent XMCD of \SI{87.5 \pm 1.5}{\percent} as compared to the saturated signal. The field-dependent Mn XMCD signal in Fig.~\ref{Magnetism_figure}(b), shows an open hysteresis loop with a coercive field of \SI{-48}{\milli\tesla} and a pronounced saturation behaviour above \SI{100}{\milli\tesla} in the FM phase. The approximate square-shape of the hysteresis indicates an out-of-plane easy axis, as further confirmed by measurements in in-plane geometry [see supplementary material, Fig.~(S7)]. Following the temperature dependence of the remnant XMCD signal in Fig.~\ref{Magnetism_figure}(d), a reduction of the remnant magnetization towards a critical temperature of $\sim$\SI{14.9}{\kelvin} is observed. The decay towards $T_c$ is well described by a critical exponent of $\beta$ = 0.484 [see supplementary material, Fig. (S8)]. 

To investigate the global magnetic properties we performed magneto-transport measurements. For these experiments thinner Bi$_2$Te$_3$ epilayers (3 QLs) were employed to minimize contributions from non-magnetic Bi$_2$Te$_3$ to the charge transport. The samples were covered in-situ, after growth, by a 30 nm thick BaF$_2$ capping layer to avoid surface contamination and oxidation during air exposure (see supplementary material for more detailed information). We contacted a macroscopic millimeter-scale as-grown thin film in a pseudo-Hall bar geometry. The inset in Fig.~\ref{Magnetism_figure}(c) shows a schematic of the geometry for Hall and longitudinal configurations, where the current is applied through the bar while measuring the resistance perpendicular and parallel to the current, respectively [see also SI, Fig. (S6)]. Fig.~\ref{Magnetism_figure}(c) displays data of Hall measurements performed with perpendicular magnetic fields at different temperatures from \SI{4}{\kelvin} to \SI{20}{\kelvin}, after subtraction of the linear normal Hall-effect background. We observe a pronounced anomalous Hall effect (AHE) hysteresis at \SI{4}{\kelvin} with an amplitude of \SI{2.8}{\ohm}. Further, at \SI{4}{\kelvin}, the coercive field reaches \SI{90}{\milli\tesla}, and decreases above \SI{10}{\kelvin} (\SI{10}{\milli\tesla}). In Fig.~\ref{Magnetism_figure}(e), the AHE hysteresis amplitude is extracted from the anomalous Hall curves. The amplitude of the hysteresis decreases with increasing temperature and the loop finally vanishes, pointing to a critical temperature of $\sim$\SI{17}{\kelvin}, comparable to the one observed in XMCD. Since the Bi$_2$Te$_3$ epilayer is non-magnetic and given the XMCD data, the AHE hysteresis can be attributed to the MnBi$_2$Te$_4$ monolayer.\\
Hence, our results from these two complementary probes establish a FM state in monolayer MnBi$_2$Te$_4$/Bi$_2$Te$_3$ with a $T_c$ of \SI{15 \pm 2}{\kelvin} and out-of-plane anisotropy [see SI, Fig. (S7)]. This refutes earlier claims of paramagnetic properties at lower temperatures \cite{fukusawa:21_absenceFMmonolayer}, but is in line with theoretical predictions \cite{otrokov2017highly} and experimental results for thin MnBi$_2$Te$_4$ flakes exfoliated from bulk crystals \cite{PhysRevX.11.011003}. Our magnetic data also vastly deviates from a recent work \cite{li:2021designerferromagnet}, where the growth of monolayer MnBi$_2$Te$_4$/Bi$_2$Te$_3$ was claimed without appropriate evidence from structural data.



We next discuss the temperature dependence of the TSS dispersion across $T_c$ of the MnBi$_2$Te$_4$ monolayer. As schematically shown in Fig.~\ref{StructureFigure}(a), an out-of-plane FM layer at the surface of a 3D TI is expected to break TRS, lifting the topological protection and allowing for a gap to open at the Dirac point. In Fig.~\ref{ARPES_Figure}(a) we present high-resolution ARPES measurements of the TSS along $k_x$ and $k_y$ for temperatures well above and below $T_c$, acquired with a photon energy of $h\nu$ = \SI{12}{\electronvolt}. To exclude angular misalignment, the ARPES intensity around the $\bar{\Gamma}$ point is sampled in small increments along both $k_{\|}$ components. The rather pronounced intensity asymmetry along $k_y$ arises from the experimental geometry, with p-polarized light incident in the $yz$ plane.

The ARPES data acquired below $T_c$ ($T$~=~\SI{1.5}{\kelvin}) reveal a reduction of spectral weight at the Dirac point, \textit{i.e.} at around \SI{-0.14}{\electronvolt}, and a splitting into two peak components at $\bar{\Gamma}$, as directly visible in the energy distribution curve (EDC) [Fig.~\ref{ARPES_Figure}(a)]. By contrast, the spectral-weight reduction and the splitting disappear above $T_c$ ($T~$=~\SI{20}{\kelvin}), demonstrating their magnetic origin. We determine an exchange splitting of $\Delta = 35$~meV at $T~$=~\SI{1.5}{\kelvin} [see supplementary information, Fig. (S12)], which corresponds to the size of the gap in the TSS and is in good agreement with theory \cite{otrokov2017highly}.

The temperature series in Fig.~\ref{ARPES_Figure}(b) shows how the spectral-weight distribution of the TSS at $\bar{\Gamma}$ evolves as the temperature is varied across $T_c$. With increasing temperature the double-peak structure gradually becomes less well-defined but remains observable in the FM regime, up to 12~K. At and above $T_c\sim15$~K, the spectra differ qualitatively from the FM regime, showing a single maximum, a rather symmetric line-shape and no discernible double-peak structure or peak shoulder. Hence, the data indicate a closing of the magnetic gap at $T_c$, as expected from our XMCD experiments that show a vanishing net magnetization $m_z$ and, thus, a globally preserved TRS. The temperature-dependent line shape in Fig.~\ref{ARPES_Figure}(b) is reasonably well captured by a simulation that assumes a power-law dependence of the magnetic gap $\Delta(T)=\Delta_0(1-T/T_c)^{\beta}$, with $\beta = 0.484$ estimated from the XMCD data [see Supplementary material Fig. (S8)]. Hence, our combined XMCD and ARPES results support an approximate relation $\Delta \propto m_z$. The origin of the gap can therefore be attributed to the FM order at the surface.

To study the momentum-dependent effect of $m_z$ on the TSS dispersion, we consider a more detailed comparison of the ARPES data near the $\bar{\Gamma}$ point above and below $T_c$ [Fig.~3(c)]. It is evident from the data that the FM order affects the dispersion only in a narrow $k_{\|}$ range of about $\pm$0.01\AA$^{-1}$. At larger momenta the dispersion remains largely unchanged. This observation is in line with an effective model for a TSS perturbed by an exchange field $m_z$ \cite{Zhang:11.10}, described by: $H=v_F(\sigma_xk_y-\sigma_yk_x)+m_z\sigma_z$. The ratio $m_z/v_F$, quantifying the relative strengths of exchange interaction and spin-orbit interaction, provides a natural scale $\delta k$ on which $m_z$ modifies the TSS dispersion and wave function. From our experimental data we may estimate $\delta k$ to $\pm$0.01\AA$^{-1}$, which indeed quantitatively matches calculations of the momentum-resolved Berry curvature in MnBi$_2$Te$_4$/Bi$_2$Te$_3$ \cite{otrokov2017highly}.


Taken collectively, our results demonstrate a 2D ferromagnetic state and a robust gap at the surface of a 3D topological insulator. Preserving the bulk topological properties, this situation differs from the previous paradigm of 3D magnetic topological insulators where time-reversal symmetry is broken in the bulk \cite{Chang2013:QAH_dopedTI,Tokura2019:MTI,otrokov:19,rienks2019large, Deng:20QAH,PhysRevX.11.011039}. The present experiments support a ferromagnetic-extension effect in epitaxial monolayer MnBi$_2$Te$_4$ on Bi$_2$Te$_3$ where the topological surface state migrates into the ferromagnetic layer, yielding a robust magnetic gap \cite{otrokov2017highly}. Our findings establish monolayer MnBi$_2$Te$_4$/Bi$_2$Te$_3$ as a versatile platform for exploring magnetic topological quantum phenomena. In particular, the van der Waals nature of both materials will facilitate integration with other magnetic or superconducting 2D crystals as a route to axion physics and chiral Majorana states. While signatures of the QAH effect and axion physics were reported in microscopic exfoliated MnBi$_2$Te$_4$ flakes, fine tuning of the layer thickness is required to achieve an uncompensated or compensated antiferromagnetic state \cite{Deng:20QAH,liu2020robust}. Instead, the ferromagnetic state and large magnetic gap in monolayer MnBi$_2$Te$_4$/Bi$_2$Te$_3$, which we report here, provide ideal prerequisites for a robust QAH state \cite{Zhang:11.10} and for tailoring the magnetic topological properties, e.g., in heterostructures with other 2D magnets \cite{he2017tailoring,gibertini2019magnetic,fu2020exchange}.

The direct measurement of surface gaps in 3D magnetic topological insulators has been notably difficult. No clear evidence of magnetic gaps could be obtained in most materials, such as bulk MnBi$_2$Te$_4$ \cite{otrokov:19,chen:19topological,hao2019gapless}, while signatures were detected in some instances for selected surface terminations \cite{rienks2019large,PhysRevX.11.011039}. Overall, the conditions under which 3D magnetic order manifests in the topological surface electronic structure remain controversial and may be influenced by disorder and surface defects \cite{Garnica:22MBTdefect} or by deviations between bulk and surface magnetism \cite{hao2019gapless}. As our data shows, the strictly 2D FM order in single monolayer MnBi$_2$Te$_4$/Bi$_2$Te$_3$ yields a large magnetic gap, which will be crucial for integration into spin-electronic devices based on 2D heterostructures.

\section{Acknowledgments}
This work is funded by the Deutsche Forschungsgemeinschaft (DFG, German Research Foundation) through Project-ID 258499086 - SFB 1170 (Project A01 and C06) and through the W\"urzburg-Dresden Cluster of Excellence on Complexity and Topology in Quantum Matter --\textit{ct.qmat} Project-ID 390858490 - EXC 2147.\\
L.V. was supported by the Leibnitz Association through the Leibniz Competition. S.L.M. acknowledges FAPESP (ID 2019/01946-1) and CNPq (ID 310432/2020-0).\\
We acknowledge ALBA for beamtime provision at beamline 29 under proposal number ID-2021035070 and ID-2021035101. We acknowledge BESSY for financial support and beamtime provision at UE112\_PGM-2b-1\^{}3 under proposal nr. 211-10192 ST and 211-11130 ST. Furthermore we thank Dr. Rui Lou for technical assistance during the experiments.

\bibliographystyle{apsrev}
\bibliography{references4}

\end{document}